\documentclass[prl, reprint, superscriptaddress]{revtex4-1}
\usepackage{hyperref}
\usepackage{dcolumn}
\usepackage{geometry}  
          
\usepackage{graphicx}	

\usepackage{amssymb}

\newcommand{\perroothz}{Hz$^{-1/2}$}
\newcommand{\anoise} [1] { {#1}m~s$^{-2}$~Hz$^{-1/2}$}

\newcommand{\epers}{e\thinspace s$^{-1}$}

\begin{document}

\title{Charge-induced force-noise on free-falling test masses: results from LISA Pathfinder}




\author{M.~Armano}
\affiliation{European Space Astronomy Centre, European Space Agency, Villanueva de la Ca\~{n}ada, 28692 Madrid, Spain}

\author{H.~Audley}
\affiliation{Albert-Einstein-Institut, Max-Planck-Institut f\"ur Gravitationsphysik und Universit\"at Hannover, 30167 Hannover, Germany}

\author{G.~Auger}
\affiliation{APC UMR7164, Universit\'e Paris Diderot, Paris, France}

\author{J.~T.~Baird}
\affiliation{High Energy Physics Group, Department of Physics, Imperial College London, Blackett Laboratory, Prince Consort Road, London SW7 2BW, UK}

\author{P.~Binetruy}
\affiliation{APC UMR7164, Universit\'e Paris Diderot, Paris, France}

\author{M.~Born}
\affiliation{Albert-Einstein-Institut, Max-Planck-Institut f\"ur Gravitationsphysik und Universit\"at Hannover, 30167 Hannover, Germany}

\author{D.~Bortoluzzi}
\affiliation{Department of Industrial Engineering, University of Trento, via Sommarive 9, 38123 Trento, and Trento Institute for Fundamental Physics and Application / INFN}

\author{N.~Brandt}
\affiliation{Airbus Defence and Space, Claude-Dornier-Strasse, 88090 Immenstaad, Germany}

\author{A.~Bursi}
\affiliation{CGS S.p.A, Compagnia Generale per lo Spazio, Via Gallarate, 150 - 20151 Milano, Italy}

\author{M.~Caleno}
\affiliation{European Space Technology Centre, European Space Agency, Keplerlaan 1, 2200 AG Noordwijk, The Netherlands}

\author{A.~Cavalleri}
\affiliation{Dipartimento di Fisica, Universit\`a di Trento and Trento Institute for Fundamental Physics and Application / INFN, 38123 Povo, Trento, Italy}

\author{A.~Cesarini}
\affiliation{Dipartimento di Fisica, Universit\`a di Trento and Trento Institute for Fundamental Physics and Application / INFN, 38123 Povo, Trento, Italy}

\author{M.~Cruise}
\affiliation{Department of Physics and Astronomy, University of Birmingham,Birmingham, Edgbaston Park Road, Birmingham, B15 2TT, UK}

\author{K.~Danzmann} 
\affiliation{Albert-Einstein-Institut, Max-Planck-Institut f\"ur Gravitationsphysik und Universit\"at Hannover, 30167 Hannover, Germany}

\author{M.~de Deus Silva}
\affiliation{European Space Astronomy Centre, European Space Agency, Villanueva de la Ca\~{n}ada, 28692 Madrid, Spain}

\author{I.~Diepholz} 
\affiliation{Albert-Einstein-Institut, Max-Planck-Institut f\"ur Gravitationsphysik und Universit\"at Hannover, 30167 Hannover, Germany}

\author{R.~Dolesi}
\affiliation{Dipartimento di Fisica, Universit\`a di Trento and Trento Institute for Fundamental Physics and Application / INFN, 38123 Povo, Trento, Italy}

\author{N.~Dunbar}
\affiliation{Airbus Defence and Space, Gunnels Wood Road, Stevenage, Hertfordshire, SG1 2AS, UK }

\author{L.~Ferraioli}
\affiliation{Institut f\"ur Geophysik, ETH Z\"urich, Sonneggstrasse 5, CH-8092, Z\"urich, Switzerland}

\author{V.~Ferroni}
\affiliation{Dipartimento di Fisica, Universit\`a di Trento and Trento Institute for Fundamental Physics and Application / INFN, 38123 Povo, Trento, Italy}

\author{E.~D.~Fitzsimons}
\affiliation{UK Astronomy Technology Centre, Royal Observatory, Edinburgh, EH9 3HJ, UK}

\author{R.~Flatscher}
\affiliation{Airbus Defence and Space, Claude-Dornier-Strasse, 88090 Immenstaad, Germany}

\author{M.~Freschi}
\affiliation{European Space Astronomy Centre, European Space Agency, Villanueva de la Ca\~{n}ada, 28692 Madrid, Spain}

\author{J.~Gallegos}
\affiliation{European Space Astronomy Centre, European Space Agency, Villanueva de la Ca\~{n}ada, 28692 Madrid, Spain}

\author{C.~Garc\'ia Marirrodriga}
\affiliation{European Space Technology Centre, European Space Agency, Keplerlaan 1, 2200 AG Noordwijk, The Netherlands}

\author{R.~Gerndt}
\affiliation{Airbus Defence and Space, Claude-Dornier-Strasse, 88090 Immenstaad, Germany}

\author{L.~Gesa}
\affiliation{Institut de Ci\`encies de l'Espai (CSIC-IEEC), Campus UAB, Carrer de Can Magrans s/n, 08193 Cerdanyola del Vall\`es, Spain}

\author{F.~Gibert}
\affiliation{Dipartimento di Fisica, Universit\`a di Trento and Trento Institute for Fundamental Physics and Application / INFN, 38123 Povo, Trento, Italy}

\author{D.~Giardini}
\affiliation{Institut f\"ur Geophysik, ETH Z\"urich, Sonneggstrasse 5, CH-8092, Z\"urich, Switzerland}

\author{R.~Giusteri}
\affiliation{Dipartimento di Fisica, Universit\`a di Trento and Trento Institute for Fundamental Physics and Application / INFN, 38123 Povo, Trento, Italy}

\author{C.~Grimani}
\affiliation{DiSPeA, Universit\`a di Urbino ``Carlo Bo", Via S. Chiara, 27 61029 Urbino/INFN, Italy}

\author{J.~Grzymisch}
\affiliation{European Space Technology Centre, European Space Agency, Keplerlaan 1, 2200 AG Noordwijk, The Netherlands}

\author{I.~Harrison}
\affiliation{European Space Operations Centre, European Space Agency, 64293 Darmstadt, Germany }

\author{G.~Heinzel} 
\affiliation{Albert-Einstein-Institut, Max-Planck-Institut f\"ur Gravitationsphysik und Universit\"at Hannover, 30167 Hannover, Germany}

\author{M.~Hewitson} 
\affiliation{Albert-Einstein-Institut, Max-Planck-Institut f\"ur Gravitationsphysik und Universit\"at Hannover, 30167 Hannover, Germany}

\author{D.~Hollington}
\affiliation{High Energy Physics Group, Department of Physics, Imperial College London, Blackett Laboratory, Prince Consort Road, London SW7 2BW, UK}

\author{M.~Hueller}
\affiliation{Dipartimento di Fisica, Universit\`a di Trento and Trento Institute for Fundamental Physics and Application / INFN, 38123 Povo, Trento, Italy}

\author{J.~Huesler}
\affiliation{European Space Technology Centre, European Space Agency, Keplerlaan 1, 2200 AG Noordwijk, The Netherlands}

\author{H.~Inchausp\'e}
\altaffiliation[Present address: ]{Physics and Instrumentation Department, ONERA, the French Aerospace Lab, 92320 Chatillon, France}
\affiliation{APC UMR7164, Universit\'e Paris Diderot, Paris, France}

\author{O.~Jennrich}
\affiliation{European Space Technology Centre, European Space Agency, Keplerlaan 1, 2200 AG Noordwijk, The Netherlands}

\author{P.~Jetzer}
\affiliation{Physik Institut, Universit\"at Z\"urich, Winterthurerstrasse 190, CH-8057 Z\"urich, Switzerland}

\author{B.~Johlander}
\affiliation{European Space Technology Centre, European Space Agency, Keplerlaan 1, 2200 AG Noordwijk, The Netherlands}

\author{N.~Karnesis}
\affiliation{Albert-Einstein-Institut, Max-Planck-Institut f\"ur Gravitationsphysik und Universit\"at Hannover, 30167 Hannover, Germany}

\author{B.~Kaune}
\affiliation{Albert-Einstein-Institut, Max-Planck-Institut f\"ur Gravitationsphysik und Universit\"at Hannover, 30167 Hannover, Germany}

\author{C.~J.~Killow}
\affiliation{SUPA, Institute for Gravitational Research, School of Physics and Astronomy, University of Glasgow, Glasgow, G12 8QQ, UK}

\author{N.~Korsakova}
\affiliation{SUPA, Institute for Gravitational Research, School of Physics and Astronomy, University of Glasgow, Glasgow, G12 8QQ, UK}

\author{I.~Lloro}
\affiliation{Institut de Ci\`encies de l'Espai (CSIC-IEEC), Campus UAB, Carrer de Can Magrans s/n, 08193 Cerdanyola del Vall\`es, Spain}

\author{L.~Liu}
\affiliation{Dipartimento di Fisica, Universit\`a di Trento and Trento Institute for Fundamental Physics and Application / INFN, 38123 Povo, Trento, Italy}

\author{R.~Maarschalkerweerd}
\affiliation{European Space Operations Centre, European Space Agency, 64293 Darmstadt, Germany }

\author{S.~Madden}
\affiliation{European Space Technology Centre, European Space Agency, Keplerlaan 1, 2200 AG Noordwijk, The Netherlands}

\author{D.~Mance}
\affiliation{Institut f\"ur Geophysik, ETH Z\"urich, Sonneggstrasse 5, CH-8092, Z\"urich, Switzerland}

\author{V.~Mart\'{i}n}
\affiliation{Institut de Ci\`encies de l'Espai (CSIC-IEEC), Campus UAB, Carrer de Can Magrans s/n, 08193 Cerdanyola del Vall\`es, Spain}

\author{L.~Martin-Polo}
\affiliation{European Space Astronomy Centre, European Space Agency, Villanueva de la Ca\~{n}ada, 28692 Madrid, Spain}

\author{J.~Martino} 
\affiliation{APC UMR7164, Universit\'e Paris Diderot, Paris, France}

\author{F.~Martin-Porqueras}
\affiliation{European Space Astronomy Centre, European Space Agency, Villanueva de la Ca\~{n}ada, 28692 Madrid, Spain}

\author{I.~Mateos}
\affiliation{Institut de Ci\`encies de l'Espai (CSIC-IEEC), Campus UAB, Carrer de Can Magrans s/n, 08193 Cerdanyola del Vall\`es, Spain}

\author{P.~W.~McNamara}
\affiliation{European Space Technology Centre, European Space Agency, Keplerlaan 1, 2200 AG Noordwijk, The Netherlands}

\author{J.~Mendes}
\affiliation{European Space Operations Centre, European Space Agency, 64293 Darmstadt, Germany }

\author{L.~Mendes}
\affiliation{European Space Astronomy Centre, European Space Agency, Villanueva de la Ca\~{n}ada, 28692 Madrid, Spain}

\author{A.~Moroni}
\affiliation{CGS S.p.A, Compagnia Generale per lo Spazio, Via Gallarate, 150 - 20151 Milano, Italy}

\author{M.~Nofrarias}
\affiliation{Institut de Ci\`encies de l'Espai (CSIC-IEEC), Campus UAB, Carrer de Can Magrans s/n, 08193 Cerdanyola del Vall\`es, Spain}

\author{S.~Paczkowski}
\affiliation{Albert-Einstein-Institut, Max-Planck-Institut f\"ur Gravitationsphysik und Universit\"at Hannover, 30167 Hannover, Germany}

\author{M.~Perreur-Lloyd}
\affiliation{SUPA, Institute for Gravitational Research, School of Physics and Astronomy, University of Glasgow, Glasgow, G12 8QQ, UK}

\author{A.~Petiteau} 
\affiliation{APC UMR7164, Universit\'e Paris Diderot, Paris, France}

\author{P.~Pivato}
\affiliation{Dipartimento di Fisica, Universit\`a di Trento and Trento Institute for Fundamental Physics and Application / INFN, 38123 Povo, Trento, Italy}

\author{E.~Plagnol} 
\affiliation{APC UMR7164, Universit\'e Paris Diderot, Paris, France}

\author{P.~Prat}
\affiliation{APC UMR7164, Universit\'e Paris Diderot, Paris, France}

\author{U.~Ragnit}
\affiliation{European Space Technology Centre, European Space Agency, Keplerlaan 1, 2200 AG Noordwijk, The Netherlands}

\author{J.~Ramos-Castro}
\affiliation{Department d'Enginyeria Electr\`onica, Universitat Polit\`ecnica de Catalunya,  08034 Barcelona, Spain}
\affiliation{Institut d'Estudis Espacials de Catalunya (IEEC), C/ Gran Capit\`a 2-4, 08034 Barcelona, Spain}

\author{J.~Reiche}
\affiliation{Albert-Einstein-Institut, Max-Planck-Institut f\"ur Gravitationsphysik und Universit\"at Hannover, 30167 Hannover, Germany}

\author{J.~A.~Romera Perez}
\affiliation{European Space Technology Centre, European Space Agency, Keplerlaan 1, 2200 AG Noordwijk, The Netherlands}

\author{D.~I.~Robertson} 
\affiliation{SUPA, Institute for Gravitational Research, School of Physics and Astronomy, University of Glasgow, Glasgow, G12 8QQ, UK}

\author{H.~Rozemeijer}
\affiliation{European Space Technology Centre, European Space Agency, Keplerlaan 1, 2200 AG Noordwijk, The Netherlands}

\author{F.~Rivas}
\affiliation{Institut de Ci\`encies de l'Espai (CSIC-IEEC), Campus UAB, Carrer de Can Magrans s/n, 08193 Cerdanyola del Vall\`es, Spain}

\author{G.~Russano}
\affiliation{Dipartimento di Fisica, Universit\`a di Trento and Trento Institute for Fundamental Physics and Application / INFN, 38123 Povo, Trento, Italy}

\author{P.~Sarra}
\affiliation{CGS S.p.A, Compagnia Generale per lo Spazio, Via Gallarate, 150 - 20151 Milano, Italy}

\author{A.~Schleicher}
\affiliation{Airbus Defence and Space, Claude-Dornier-Strasse, 88090 Immenstaad, Germany}

\author{J.~Slutsky}
\affiliation{NASA Goddard Space Flight Center, 8800 Greenbelt Road, Greenbelt, MD 20771, USA}

\author{C.~Sopuerta}
\affiliation{Institut de Ci\`encies de l'Espai (CSIC-IEEC), Campus UAB, Carrer de Can Magrans s/n, 08193 Cerdanyola del Vall\`es, Spain}

\author{T.~J.~Sumner} 
\affiliation{High Energy Physics Group, Department of Physics, Imperial College London, Blackett Laboratory, Prince Consort Road, London SW7 2BW, UK}

\author{D.~Texier}
\affiliation{European Space Astronomy Centre, European Space Agency, Villanueva de la Ca\~{n}ada, 28692 Madrid, Spain}

\author{J.~I.~Thorpe}
\affiliation{NASA Goddard Space Flight Center, 8800 Greenbelt Road, Greenbelt, MD 20771, USA}

\author{C.~Trenkel}
\affiliation{Airbus Defence and Space, Gunnels Wood Road, Stevenage, Hertfordshire, SG1 2AS, UK }

\author{D.~Vetrugno}
\affiliation{Dipartimento di Fisica, Universit\`a di Trento and Trento Institute for Fundamental Physics and Application / INFN, 38123 Povo, Trento, Italy}

\author{S.~Vitale}
\affiliation{Dipartimento di Fisica, Universit\`a di Trento and Trento Institute for Fundamental Physics and Application / INFN, 38123 Povo, Trento, Italy}

\author{G.~Wanner} 
\affiliation{Albert-Einstein-Institut, Max-Planck-Institut f\"ur Gravitationsphysik und Universit\"at Hannover, 30167 Hannover, Germany}

\author{H.~Ward}
\affiliation{SUPA, Institute for Gravitational Research, School of Physics and Astronomy, University of Glasgow, Glasgow, G12 8QQ, UK}

\author{P.~J.~Wass} 
\affiliation{High Energy Physics Group, Department of Physics, Imperial College London, Blackett Laboratory, Prince Consort Road, London SW7 2BW, UK}

\author{D.~Wealthy}
\affiliation{Airbus Defence and Space, Gunnels Wood Road, Stevenage, Hertfordshire, SG1 2AS, UK }

\author{W.~J.~Weber}
\affiliation{Dipartimento di Fisica, Universit\`a di Trento and Trento Institute for Fundamental Physics and Application / INFN, 38123 Povo, Trento, Italy}

\author{A.~Wittchen}
\affiliation{Albert-Einstein-Institut, Max-Planck-Institut f\"ur Gravitationsphysik und Universit\"at Hannover, 30167 Hannover, Germany}

\author{C.~Zanoni}
\affiliation{Department of Industrial Engineering, University of Trento, via Sommarive 9, 38123 Trento, and Trento Institute for Fundamental Physics and Application / INFN}

\author{T.~Ziegler}
\affiliation{Airbus Defence and Space, Claude-Dornier-Strasse, 88090 Immenstaad, Germany}

\author{P.~Zweifel}
\affiliation{Institut f\"ur Geophysik, ETH Z\"urich, Sonneggstrasse 5, CH-8092, Z\"urich, Switzerland}

\collaboration{The LISA Pathfinder Collaboration}
\noaffiliation

\begin{abstract}

We report on electrostatic measurements made on board the European Space Agency mission LISA Pathfinder. Detailed measurements of the charge-induced electrostatic forces exerted on free-falling test masses (TMs) inside the capacitive gravitational reference sensor are the first made in a relevant environment for a space-based gravitational wave detector. Employing a combination of charge control and electric-field compensation, we show that the level of charge-induced acceleration noise on a single TM can be maintained at a level close to 1.0~\anoise{f} across the 0.1-100~mHz frequency band that is crucial to an observatory such as LISA. Using dedicated measurements that detect these effects in the differential acceleration between the two test masses, we resolve the stochastic nature of the TM charge build up due to interplanetary cosmic rays and the TM charge-to-force coupling through stray electric fields in the sensor. All our measurements are in good agreement with predictions based on a relatively simple electrostatic model of the LISA Pathfinder instrument. 

\end{abstract}

\maketitle


Sensitive gravitational experiments employ quasi-free-falling isolated test masses as a reference system for the measurement of the local curvature of space-time. Electrostatic free charge and stray potentials introduce unwanted disturbances that can limit measurement precision. The effect is relevant for gravitational wave observatories both in space \cite{Shaul2005, Antonucci} and on-ground \cite{Martynov2016}, tests of the equivalence principal \cite{Sumner} and measurements of relativistic effects on precessing gyroscopes \cite{Buchman2011}. 

The LISA Pathfinder spacecraft \cite{Mcnamara2008}, a technology-demonstration experiment for a space-based gravitational wave observatory, LISA \cite{LisaYB2011}, was launched on December 3 2015. The aim of the mission was to demonstrate the ability to fly free-falling test masses in a single spacecraft with a differential acceleration noise below 30~\anoise{f}. The sensitivity of the instrument has far-exceeded its design specification, achieving a level close to the LISA goal from 0.1-100~mHz and around 5~\anoise{f} in the mHz band \cite{Armano2016}. In this paper we describe the measurements and techniques used to minimise charge-related electrostatic forces and evaluate their contribution to the differential acceleration noise of the test masses.

The LISA Pathfinder test masses, identical to those for LISA, are 46-mm cubes of mass 1.928~kg made from a gold-platinum alloy. They sit within a 6 degree-of-freedom capacitive position sensor and actuator, the gravitational reference sensor (GRS) \cite{Dolesi2003, Weber2003}. The masses are separated from the walls of the sensor by gaps of between 2.9 and 4~mm and have no grounding wire. All test mass and sensor surfaces are gold-coated. The large gaps mitigate the impact of surface forces and the absence of a grounding wire eliminates thermal noise associated with mechanical damping that dominates the low-frequency performance of accelerometers in existing geodesy and fundamental physics missions. 

The achieved level of sensitivity to the differential acceleration of the test masses level is made possible by an additional high-precision readout along the $x$-axis provided by a laser interferometer \cite{Heinzel2003, Heinzel2004, Audley2011} with a readout noise of 35~fm\thinspace Hz$^{-1/2}$ measured above 60~mHz \cite{Armano2016}. 

The GRS consists of a system of 12 electrodes for TM position sensing and actuation, and a further 6 for capacitive biasing of the test mass at 100~kHz. Actuation is achieved with audio-frequency sinusoidal voltages. DC or slowly-varying ($f$$\sim$mHz) voltage signals can be applied to measure TM charge and balance stray electrostatic fields, as will be discussed shortly. Voltages on all electrodes originate from the GRS front-end electronics \cite{Mance2012}. 

High-energy cosmic rays and solar energetic particles, mostly protons, penetrate the spacecraft and instrument shielding depositing charge on the test mass, either by stopping directly or by secondary emission \cite{Jafry1996, Araujo2005, Wass2005, Grimani2005}. Limiting charge accumulation on the electrically isolated TMs is needed to control electrostatic forces, the subject of this paper.Ê In LISA Pathfinder, non-contact discharge is achieved by illuminating the sensor and test-mass surfaces with UV light and transferring charge by photoemission \cite{Schulte2006} in a similar way to that already demonstrated on GP-B \cite{Buchman1995}. A detailed account of the performance of the LISA Pathfinder UV discharge system will be provided in a subsequent article.



As well as providing desired actuation forces, the GRS is a source of unwanted electrostatic disturbances on the test mass \cite{Weber2003, Araujo2003}. With the TM centered, the dominant source of electrostatic force noise is the interaction between the TM charge, $q$, and stray electric fields represented by an effective potential difference between opposite sides of the TM, $\Delta_x$ \footnote{$\Delta_x$ is the equivalent uniform single GRS $x$-electrode potential that would give the same average stray field along the $x$ axis.}. The resulting force along the $x$-axis, following the notation of \cite{Antonucci}, is
\begin{equation}\label{Fx}
F_x(q) = -\frac{q}{C_{\textsc{t}}} \left| \frac{\partial{C_x}}{\partial{x}} \right| \Delta_x,
\end{equation}
where $\frac{\partial{C_x}}{\partial{x}}$ is the derivative of a single sensing electrode capacitance with respect to TM displacement along $x$ and $C_{\textsc{t}}$ is the total capacitance of the test mass with respect to the GRS \footnote{Finite element modelling calculates $C_\textsc{t}$= 34.2~pF and $\frac{\partial{C_x}}{\partial{x}}$= 291~pF\thinspace m$^{-1}$.}. %


In LISA Pathfinder, the principle science observable is the differential force per unit mass acting on the two TMs, $\Delta g \equiv \frac{F_{2x}}{m_2} - \frac{F_{1x}}{m_1}$. The measurement is thus sensitive to the in-band fluctuations of both $\Delta_x$ and $q$ for the two TM. Force noise is produced by a non-zero charge $q$, coupling with fluctuations in the average potential difference $\Delta_x$ and, likewise, stochastic charge fluctuations mixing with any non-zero potential difference. We measure these effects with a number of dedicated techniques.

The test-mass charge, $q$ can be detected in its effect on the TM potential,  $\delta V_{TM} = \frac{\delta q}{C_{T}}$, measured  by applying sinusoidally varying voltages with amplitude $V_{\textsc{mod}}$ and frequency, $f_{\textsc{mod}}$ on the $x$-axis electrodes, a technique well-demonstrated in ground-based investigations \cite{Wass2006, Pollack2010}. The resulting force on the TM is $F_x(f_{\textsc{mod}}) = -4 \left| \frac{\partial{C_x}}{\partial{x}}\right| V_{\textsc{mod}}V_{\textsc{tm}}$. A continuous measurement provides an extended time series of $q\left(t\right)$ from which the low-frequency behavior of the charge build up can be studied.  

To measure the relevant stray potential-difference $\Delta_x$, we introduce a potential $\pm V_{\textsc{comp}}$ to each $x$ electrode. Following the method described in \cite{Antonucci} it is possible to estimate $\frac{\partial{F_x}}{\partial{q}}$ as a function of $V_{\textsc{comp}}$ measuring the change in $\Delta g$ as the charge of one test mass is increased in steps by photoemission under UV illumination. By choosing a value for $V_{\textsc{comp}}$ that provides an equal and opposite potential difference to $\Delta_x$, we can cancel $\frac{d F_x}{dq}$ to first order. 

The LISA Pathfinder sensitivity is sufficient that the effects of in-band fluctuations of $q$ and $\Delta_x$, described by their power spectral densities (PSDs), $S_q$ and $S_{\Delta_x}$ are measurable directly in $\Delta g$ by exaggerating $\Delta_x$ or $q$ respectively.


Simulations and ground-based laboratory measurements provide indications of the expected behavior of the test mass charge and stray potentials. High-energy physics simulations \cite{Araujo2005, Wass2005} predict that a net positive charging rate of 40-70 elementary charges per second (\epers) from galactic cosmic rays (GCR) at the minimum of the 11-year solar activity cycle (20-40~\epers{} at maximum when GCR flux is suppressed) with a charge-noise equivalent to that produced by a rate of single charges, $\lambda_{\textrm{\scriptsize{eff}}}$, of 200-400~s$^{-1}$. The amplitude spectral density (ASD) of the test-mass charge is expected to have the form $S_q^{1/2}$=$\frac{e\sqrt{2\lambda_{\text{eff}}}}{2\pi f}$ with an amplitude of 0.6-0.7~fC\thinspace \perroothz{} at 1~mHz. 

\begin{figure*}
 \includegraphics[width=\linewidth]{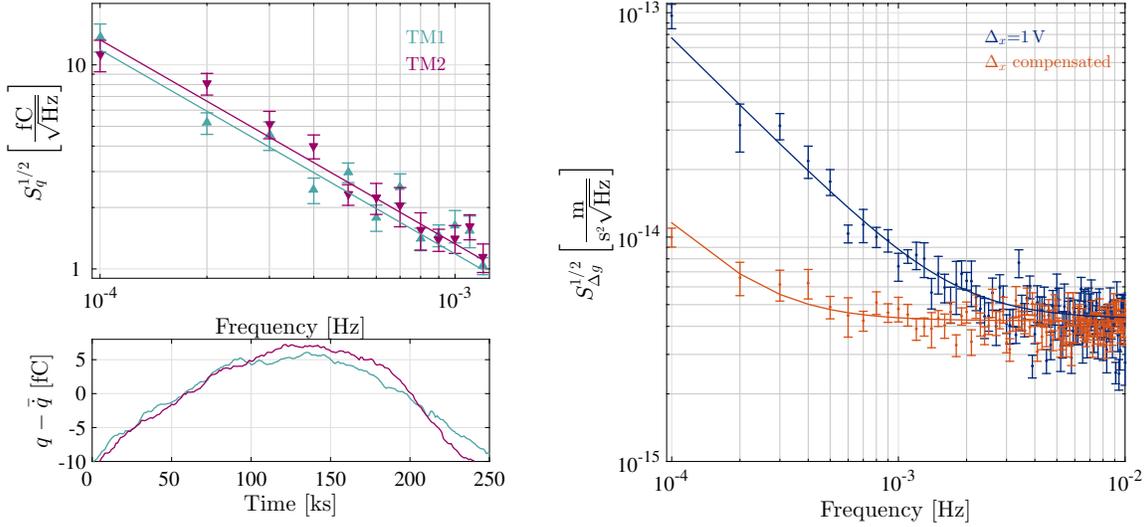} 
 \caption{Measurements of TM charge fluctuations. Upper-left: the ASD of the charge on TM\thinspace 1 ($\bigtriangleup$) and TM\thinspace 2 ($\bigtriangledown$) in the LISA band with $1/f$ fits. Lower-left: the charge time series after removal of the linear trend due to the average charge rate over the course of the 3-day measurement. Right: consecutive measurements of the ASD of $\Delta g$ with exaggerated $\Delta_x$ (dark-blue) and with $\Delta_x$ compensated to $\lesssim$3~mV (red). Continuous curves show the result of a combined fit to the background noise and $\Delta_x$-dependent $1/f$ excess.}
 \label{chargeNoise}
\end{figure*}

$\Delta_x$ originates both from surface patch potentials within the sensor and the GRS electronics. Measurements with representative systems in laboratory tests have found static levels of up to $~$100~mV \cite{Carbone2003, Weber2007, Antonucci, Pollack2008}. Tests before launch showed $S_{\Delta_x}^{1/2}$ coming from voltage fluctuations from the spacecraft electronics to be 30~$\mu$V\thinspace \perroothz{} at 1~mHz \cite{Praplan2009}. Torsion pendulum measurements using a representative TM and GRS and similar electronics have placed 2-$\sigma$ upper-limits on the total fluctuations, including patch potentials of 80~$\mu$V\thinspace \perroothz{} at 1~mHz  and 290~$\mu$V\thinspace \perroothz{} at 0.1~mHz \cite{Antonucci}. 



A $\sim$3-day measurement of $q$ was made injecting $V_{\textsc{mod}}$=3~V at $f_{\textsc{mod}}$=6 and 9~mHz on TM\thinspace 1 and TM\thinspace 2 respectively. The charge was calculated by heterodyne demodulation of $\Delta g$, with the applied $V_{\textsc{mod}}$ as the phase reference. 

The average charging rates were +22.9~\epers{} and +24.5~\epers{} on TM\thinspace 1 and 2 respectively. Over 10000-s periods, the charge rate is observed to vary by $\pm$2~\epers, caused by a combination of low frequency noise and drift. The fluctuations around the mean charging rates are shown in the lower-left panel of Figure \ref{chargeNoise}; two $>5\sigma$ glitches in the TM\thinspace 2 charge fluctuations have been removed. $S_q$ was calculated with the Welch method, averaging 11 detrended, 40000-s Blackman-Harris (BH) spectral windows with 50\%-overlap. A $f^{-2}$ fit was applied to the PSD, down-sampled by a factor 4 to remove data correlated by spectral windowing. The resulting ASD are shown in the upper left panel of Figure \ref{chargeNoise} and a summary of the results is given in Table \ref{qNoiseTab}. 

The $f^{-2}$ dependence of the charge PSD and observed absence of correlation in the two charge time-series are consistent with the model of independent Poissonian processes for the two TMs, at least down to 0.1~mHz. A common drift in the charge rate at very-low frequency (visible in the time series as a quadratic dependence after removal of the linear trends due to the average charging rates) correlates well with measurements in the on-board particle monitor and is therefore likely caused by changes in the incident particle flux. The measured charge-noise levels have roughly 5 times the expected noise power, with effective charge rates between 1000 and 1400~s$^{-1}$. Possible causes for an excess are a larger-than-expected number of high-multiplicity charging events produced by very-high energy ($\sim$TeV) cosmic rays, or a large population of low-energy ($\sim$eV) secondaries emitted from TM and GRS surfaces. These two energy regimes are the source of most uncertainty in the charging predictions \cite{Araujo2005}. 

In this measurement, made some 3-4 years before solar minimum, we find test-mass charging rates within the expected range but measurably different on the two TMs. The difference in the charge rates may originate in the different Volt-scale AC electrostatic fields used for force actuation in the two GRS. If confirmed, this would favor secondary electrons as the source of excess noise. Further measurements characterizing the charge-rate behavior in detail will be the subject of future work. 

\begin{table}
\centering
\caption{Test mass charging properties}
\label{qNoiseTab}
\begin{ruledtabular}
\begin{tabular}{c D{,}{\pm}{-1} D{,}{\pm}{-1} l}

 					&  \multicolumn{1}{c}{TM1} 		&  \multicolumn{1}{c}{TM2} 	& \\
\hline
$\dot{q}$ 				&  \multicolumn{1}{c}{+22.9} 		&  \multicolumn{1}{c}{+24.5} 	&  \epers \\
$\lambda_{\textrm{\scriptsize{eff}}}$ 	& 1060,90						& 1360,130 				& s$^{-1}$ \\
\rule{0pt}{3ex}
$\lambda_{\textrm{\scriptsize{eff(1+2)}}}$\footnote{Determined from fit to $\Delta g$ with $\Delta_x$=1V} 	&  \multicolumn{2}{c}{2200$\pm$260}				& s$^{-1}$ \\
\end{tabular}
\end {ruledtabular}
\end{table}
\begin{figure*}

 \includegraphics[width=\linewidth]{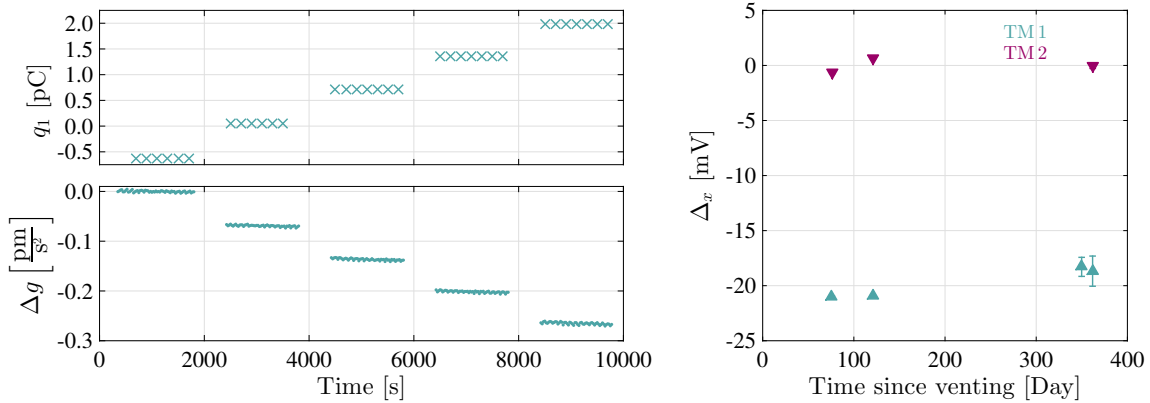} 
 \caption{Estimation of $\Delta_x$. Left: time-series of charge steps in $q_1$ ($\times$) and $\Delta g$ ($\bullet$) for measurement on TM\thinspace 1 on day 110 of 2016 with no applied compensation. Data during UV Illumination periods lasting $\sim$100~s have been removed. Right: Seven measurements of uncompensated $\Delta_x$ on TM\thinspace 1 ($\bigtriangleup$) and TM\thinspace 2 ($\bigtriangledown$), plotted against the time elapsed since opening the vacuum chambers containing the GRS to space.}
\label{dDgdq}
\end{figure*}

The spectral density of the charge noise can also be determined from a measurement of $\Delta g$ with an exaggerated potential difference $\Delta_x$. The right panel of Figure \ref{chargeNoise} shows two measurements of the ASD of $\Delta g$ calculated with the same method described for $S_q$. The first lasting $\sim$2.5-days with $\Delta_{x_1}$=$\Delta_{x_2}$=1~V and calculated by averaging 9 overlapping, 40000-s BH windows. The second $\sim$1-day later with both $\Delta_x$ compensated to $\lesssim$3~mV (as described below) using 15 windows covering nearly 4-days. We perform a combined fit to the spectra in the frequency range 0.1$\leq$$f$$\leq$20~mHz assuming a stationary background and an excess due to random charging proportional to $\Delta_x$. We find the excess noise in $\Delta g$ in the presence of the applied electric field is compatible with a total effective charge rate $\lambda_{\text{\scriptsize{eff}}_1} + \lambda_{\text{\scriptsize{eff}}_2}$=2220$\pm$260~s$^{-1}$
in good agreement with the dedicated measurement of the charge fluctuations on each test mass shown in Table \ref{qNoiseTab}. The charge noise observed in these two measurements separated by 60-days is stationary to better than 10\% and shows no measurable departure from a pure Poissonian behavior. 


In order to calculate $\Delta_x$ and the required compensation voltages, $\frac{dF}{dq}$ was determined from $\Delta g$ using four charge steps of $\sim$0.6~pC. The charge was measured throughout with $V_{\textsc{mod}}$=50~mV and $f_{\textsc{mod}}$=5~mHz. 
Figure \ref{dDgdq} shows $q$ and $\Delta g$ as a function of time through one of these measurements.
The charge steps were repeated with $V_{\textsc{comp}}$ of --20, 0, +20~mV and the dependence of $\frac{dF}{dq}$ on $V_{\textsc{comp}}$ confirms our electrostatic model to better than 2\%. Two measurements on each GRS were made 45-days apart, the second with $\Delta_x$ on TM\thinspace 1 compensated within 3~mV. At this level, the contribution to $S_{\Delta g}^{1/2}$ from random charging is 0.2~\anoise{f} at 0.1~mHz. A further three measurements were made 7~months later. The first on TM\thinspace 1 and a final measurement on each TM after reducing the temperature of the sensor from 22 to 11$^{\circ}$C.

The calculated values for $\Delta_x$, corrected for applied compensation, are given in Table \ref{Dx} and plotted in Figure \ref{dDgdq} against the system pumping time. We note that the rotational stray-field imbalance, $\Delta_{\phi}$ and $\Delta_{\eta}$, that couple TM charge into torque in analogous fashion to Eqn \ref{Fx}, have been measured in the same experiments to be roughly --32 and +36~mV for TM\thinspace 1 and +119 and +84~mV for TM\thinspace 2.  When considered with the uncompensated values measured for $\Delta _x$, roughly --20 mV and 0~mV for TM\thinspace 1 and TM\thinspace 2, the stray fields in the GRS would seem to be similar in magnitude to those observed in various measurements on GRS prototype hardware on ground \cite{Carbone2003, Weber2007, Antonucci}. Small but significant changes in $\Delta_x$ are observed for the two TMs, consistent with drifts of slightly less than mV/month, roughly an order of magnitude below typical drift values observed, for a limited number of samples, on ground \cite{Antonucci, Pollack2008}. This suggests that only very infrequent repetition of the measurement and compensation scheme will be necessary in LISA to keep acceleration noise from TM charge fluctuations below a tolerable level.  

\begin{table}
\centering
\caption{Estimates of uncompensated $\Delta_x$}
\label{Dx}
\begin{ruledtabular}
\begin{tabular}{c D{,}{\pm}{-1} D{,}{\pm}{-1} l}

Date 			& \multicolumn{1}{c}{$\Delta_{x1}$}	& \multicolumn{1}{c}{$\Delta_{x2}$} & \\
\hline
\textrm{2016-110}   	& -21.02,0.07 					& -0.67,0.07 				& mV \\
\textrm{2016-155}   	& -20.93,0.04 					& +0.66,0.03				& mV \\
\textrm{2017-018}   	& -18.3,0.9 					& \multicolumn{1}{c}{---}		& mV \\
\textrm{2017-030}   	& -18.7,1.4 					& -0.1,0.2					& mV \\

\end {tabular}
\end{ruledtabular}
\end{table}

\begin{figure}
 \includegraphics[width=\linewidth]{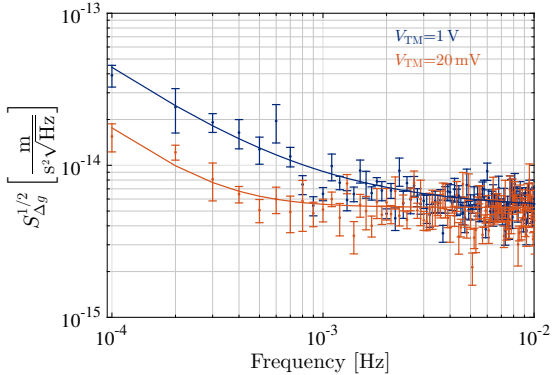} 
 \caption{Spectral density of consecutive measurements of $\Delta g$ with $V_{\textsc{tm}}$=1~V and 20~mV. A combined fit to the background and $V_{\textsc{tm}}$-dependent component is shown with continuous curves.}
 \label{specFits}
\end{figure}

The spectral density of stray voltage fluctuations was measured by increasing the test-mass charge and using a similar method to that used to observe the charge noise effect on $S^{1/2}_{\Delta g}$. Figure \ref{specFits} shows two measurements of the ASD of $\Delta g$. The first over nearly 2-days (average of 6, overlapping, 40000-s windows) with normal levels of TM potential: $\langle V_{\textsc{tm}1}\rangle$=--16~mV and $\langle V_{\textsc{tm}2}\rangle $=--24~mV. followed within a day by a measurement of just over 2-days (8, 40000-s windows) at $\langle V_{\textsc{tm}1}\rangle$=--1066~mV and $\langle V_{\textsc{tm}2}\rangle $=--1058~mV.
Fitting a polynomial to the excess in the PSD of $\Delta g$ proportional to $V_{\textsc{tm}}^2$ of the form $S_{\Delta g}$=$\left[A^2 \left(\frac{1\textrm{mHz}}{f}\right)^2 + B^2 \left(\frac{1\textrm{mHz}}{f}\right) \right] \left(\frac{V_{\textsc{tm}}}{\textrm{1V}}\right)^2$ we find $A$=3.5$\pm$0.7~\anoise{f} and $B$=6.4$\pm$0.5~\anoise{f} ($\chi^2$=1.2). This converts to an ASD of the fluctuations in $\Delta_x$ of 34$\pm$2~$\mu$V\thinspace \perroothz{} at 1~mHz and 190$\pm$30~$\mu$V\thinspace \perroothz{} at 0.1~mHz. The measured fluctuations in $\Delta_x$ are thus clearly resolved and are consistent with the upper limits placed for a GRS prototype on ground  \cite{Antonucci}.Ê They are also consistent with ground measurements of the low-frequency actuation-circuitry voltage noise, which is indistinguishable from stray surface potential fluctuations in our acceleration noise measurement.

Throughout the majority of the operational phase of the mission, $V_{\textsc{tm}}$ has been controlled within $\pm$80~mV of zero by discharging under UV illumination during weekly or fortnightly interruptions to science measurements for orbital correction. The RMS $V_{\textsc{tm}}$ in a two-week measurement period such as that described in \cite{Armano2016} is typically $<$40~mV and the contribution to $S_{\Delta g}^{1/2}$ at 0.1mHz is $1.6\pm0.2$~\anoise{f}. 


We have presented the most sensitive measurements of charge-related electrostatic forces on free-falling test masses relevant for sensitive gravitational experiments in space. Technology and mitigation methods developed for minimizing these forces on test masses in capacitive position sensors have been demonstrated and are directly transferable to LISA. Their contribution to the acceleration noise of individual test masses has been reduced to a level of roughly 1~\anoise{f} at 0.1~mHz, compatible with the budgeted allocation for electrostatic forces contributing to the total acceleration noise in LISA \cite{LisaYB2011}.  

The optimal method of test-mass charge control in a future space-based GW observatory will minimize the disruption to observations and the total contribution to acceleration noise. A continuous UV discharge scheme trades additional random-charging noise for a reduced coupling between $V_{\textsc{tm}}$ and fluctuating stray potentials. Since the latter dominates the charge-related noise in the periodic discharge scenario described in this paper by a factor of nearly 50 in power, it is likely some improvement can be obtained with a continuous scheme. This will be studied in a future experiment. 

The methods and technology described here can enable a new generation of instrumentation for gravity-gradiometry and fundamental physics with improved performance in the milli-Hz regime. 


This work has been made possible by the LISA Pathfinder mission, which is part of the space-science program of the European Space Agency. 

The French contribution has been supported by CNES (Accord Specific de projet CNES 1316634/CNRS 103747), the CNRS, the Observatoire de Paris and the University Paris-Diderot. E.~P. and H.~I. would also like to acknowledge the financial support of the UnivEarthS Labex program at Sorbonne Paris CitŽ (ANR-10-LABX-0023 and ANR-11-IDEX-0005-02). 

The Albert-Einstein-Institut acknowledges the support of the German Space Agency, DLR. The work is supported by the Federal Ministry for Economic Affairs and Energy based on a resolution of the German Bundestag (FKZ 50OQ0501 and FKZ 50OQ1601). 

The Italian contribution has been supported by Agenzia Spaziale Italiana and Instituto Nazionale di Fisica Nucleare. 

The Spanish contribution has been supported by Contracts No. AYA2010-15709 (MICINN), No. ESP2013-47637-P, and No. ESP2015-67234-P (MINECO). M.~N. acknowledges support from Fundacion General CSIC Programa ComFuturo). F.~R. acknowledges support from a Formaci—n de Personal Investigador (MINECO) contract. 

The Swiss contribution acknowledges the support of the Swiss Space Office (SSO) via the PRODEX Programme of ESA. L.~F. acknowledges the support of the Swiss National Science Foundation.

The UK groups wish to acknowledge support from the United Kingdom Space Agency (UKSA), the University of Glasgow, the University of Birmingham, Imperial College London, and the Scottish Universities Physics Alliance (SUPA). 

J.~I.~T. and J.~S. acknowledge the support of the U.S. National Aeronautics and Space Administration (NASA).

\bibliographystyle{./unsrt2}
\bibliography{LPF_Electrostatics}

\begin{thebibliography}{10}

\bibitem{Shaul2005}
D.~N.~A. Shaul, H.~M. Ara{\'{u}}jo, G.~K. Rochester, T.~J. Sumner, and P.~J.
  Wass.
\newblock {Evaluation of disturbances due to test mass charging for LISA}.
\newblock {\em Classical and Quantum Gravity}, 22(10):S297--S309, 2005.

\bibitem{Antonucci}
F.~Antonucci, A.~Cavalleri, R.~Dolesi, et~al.
\newblock {Interaction between Stray Electrostatic Fields and a Charged
  Free-Falling Test Mass}.
\newblock {\em Physical Review Letters}, 108(118):1--5, 2012.

\bibitem{Martynov2016}
D.~V. Martynov, E.~D. Hall, B.~P. Abbott, et~al.
\newblock {Sensitivity of the Advanced LIGO detectors at the beginning of
  gravitational wave astronomy}.
\newblock {\em Physical Review D}, 93(11):112004, 2016.

\bibitem{Sumner}
T.~J. Sumner, J.~Anderson, J.-P. Blaser, et~al.
\newblock {STEP (satellite test of the equivalence principle)}.
\newblock {\em Advances in Space Research}, 39:254--258, 2007.

\bibitem{Buchman2011}
S.~Buchman and J.~P. Turneaure.
\newblock {The effects of patch-potentials on the gravity probe B gyroscopes}.
\newblock {\em Review of Scientific Instruments}, 82(7):074502, 2011.

\bibitem{Mcnamara2008}
P.~McNamara, S.~Vitale, and K.~Danzmann.
\newblock {LISA Pathfinder}.
\newblock {\em Classical and Quantum Gravity}, 25(11):114034, 2008.

\bibitem{LisaYB2011}
K.~Danzmann, T.~A. Prince, P.~Binetruy, et~al.
\newblock {LISA: Revealing a hidden Universe}.
\newblock {\em ESA Assessment Study Report}, ESA/SRE(2011):3.

\bibitem{Armano2016}
M.~Armano, H.~Audley, G.~Auger, et~al.
\newblock {Sub-Femto-g Free Fall for Space-Based Gravitational Wave
  Observatories: LISA Pathfinder Results}.
\newblock {\em Physical Review Letters}, 116(23):1--10, 2016.

\bibitem{Dolesi2003}
R.~Dolesi, D.~Bortoluzzi, P.~Bosetti, et~al.
\newblock {Gravitational sensor for LISA and its technology demonstration
  mission}.
\newblock {\em Classical and Quantum Gravity}, 20(10):S99--S108, 2003.

\bibitem{Weber2003}
W.~J. Weber, D.~Bortoluzzi, C.~Carbone, et~al.
\newblock {Position sensors for flight testing of LISA drag-free control}.
\newblock In {\em Proceedings of SPIE}, pages 31--42. SPIE, 2003.

\bibitem{Heinzel2003}
G.~Heinzel, C.~Braxmaier, R.~Schilling, et~al.
\newblock {Interferometry for the LISA technology package (LTP) aboard
  SMART-2}.
\newblock {\em Classical and Quantum Gravity}, 20(10):S153--S161, 2003.

\bibitem{Heinzel2004}
G.~Heinzel, V.~Wand, A.~Garc{\'{i}}a, et~al.
\newblock {The LTP interferometer and phasemeter}.
\newblock {\em Classical and Quantum Gravity}, 21(5):S581--S587, 2004.

\bibitem{Audley2011}
H.~Audley, K.~Danzmann, A.~Garc{\'{i}}a-Mar{\'{i}}n, et~al.
\newblock {The LISA Pathfinder interferometry—hardware and system testing}.
\newblock {\em Classical and Quantum Gravity}, 28(9):094003, 2011.

\bibitem{Mance2012}
D.~Mance.
\newblock {\em {Development of Electronic System for Sensing and Actuation of
  Test Mass of the Inertial Sensor LISA}}.
\newblock PhD thesis, University of Split, 2012.

\bibitem{Jafry1996}
Y.~Jafry, T.~J. Sumner, and S.~Buchman.
\newblock {Electrostatic charging of space-borne test bodies used in precision
  experiments}.
\newblock {\em Classical and Quantum Gravity}, 13(11):97, 1996.

\bibitem{Araujo2005}
H.~M. Ara{\'{u}}jo, P.~J. Wass, D.~N.~A. Shaul, G.~Rochester, and T.~J. Sumner.
\newblock {Detailed calculation of test-mass charging in the LISA mission}.
\newblock {\em Astroparticle Physics}, 22(5-6):451--469, 2005.

\bibitem{Wass2005}
P.~J. Wass, H.~M. Ara{\'{u}}jo, D.~N.~A. Shaul, and T.~J. Sumner.
\newblock {Test-mass charging simulations for the LISA Pathfinder mission}.
\newblock {\em Classical and Quantum Gravity}, 22(10):S311--S317, 2005.

\bibitem{Grimani2005}
C.~Grimani, H.~Vocca, G.~Bagni, et~al.
\newblock {LISA test-mass charging process due to cosmic-ray nuclei and
  electrons}.
\newblock {\em Classical and Quantum Gravity}, 22(10), 2005.

\bibitem{Schulte2006}
M.~Schulte, G.~K. Rochester, D.~N.~A. Shaul, et~al.
\newblock {The Charge-Management System on LISA-Pathfinder: Status {\&} Outlook
  for LISA}.
\newblock In {\em AIP Conference Proceedings}, volume 873, pages 165--171. AIP,
  2006.

\bibitem{Buchman1995}
S.~Buchman, T.~Quinn, G.~M. Keiser, D.~Gill, and T.~J. Sumner.
\newblock {Charge measurement and control for the Gravity Probe B gyroscopes}.
\newblock {\em Review of Scientific Instruments}, 66(1):120, 1995.

\bibitem{Araujo2003}
H~M Ara{\'{u}}jo, A.~Howard, D.~Davidge, and T.~J. Sumner.
\newblock {Charging of isolated proof masses in satellite experiments such as
  LISA}.
\newblock In {\em Proceedings of SPIE}, volume 4856, page~55. SPIE, 2003.

\bibitem{Note1}
$\Delta _x$ is the equivalent uniform single GRS $x$-electrode potential that
  would give the same average stray field along the $x$ axis.

\bibitem{Note2}
Finite element modelling calculates $C_\protect \textsc {t}$= 34.2~pF and
  ${\begingroup \partial {C_x}\endgroup \over \partial {x}}$= 291~pF\kern
  .16667em m$^{-1}$.

\bibitem{Wass2006}
P.~J. Wass, L.~Carbone, A.~Cavalleri, et~al.
\newblock {Testing of the UV discharge system for LISA Pathfinder}.
\newblock In {\em AIP Conference Proceedings}, volume 873, pages 220--224. AIP,
  2006.

\bibitem{Pollack2010}
S.~E. Pollack, M.~D. Turner, and S.~Schlamminger.
\newblock {Charge management for gravitational-wave observatories using UV
  LEDs}.
\newblock {\em Physical Review D}, 81(021101):2--6, 2010.

\bibitem{Carbone2003}
L.~Carbone, A.~Cavalleri, R.~Dolesi, et~al.
\newblock {Achieving Geodetic Motion for LISA Test Masses: Ground Testing
  Results}.
\newblock {\em Physical Review Letters}, 91(15):2--5, 2003.

\bibitem{Weber2007}
W.~J. Weber, L.~Carbone, A.~Cavalleri, et~al.
\newblock {Possibilities for measurement and compensation of stray DC electric
  fields acting on drag-free test masses}.
\newblock {\em Advances in Space Research}, 39(2):213--218, 2007.

\bibitem{Pollack2008}
S.~E. Pollack, S.~Schlamminger, and J.~H. Gundlach.
\newblock {Temporal Extent of Surface Potentials between Closely Spaced
  Metals}.
\newblock {\em Physical Review Letters}, 101(7):1--4, 2008.

\bibitem{Praplan2009}
C.~Praplan.
\newblock {S2-HEV-RP-3042}.
\newblock Technical Report 1.1, 2009.

\end{thebibliography}

\end{document}